\documentclass{aa} 
\usepackage{graphicx}
\usepackage{color}
\usepackage{txfonts}
\usepackage{natbib}
\usepackage{soul}

\input epsf.sty

\begin{document}

\title{Quasar Main Sequence: a line or a plane?}

\author{Conor Wildy \inst{1}
        \and
      Bo\.zena Czerny\inst{1}
      \and
      Swayamtrupta Panda\inst{1,2}
        }

 \offprints{wildy@cft.edu.pl}

\institute{Center for Theoretical Physics, Polish Academy of Sciences, Al. Lotnik\' ow 32/46, 02-668 Warsaw, Poland\\
     \and{Copernicus Astronomical Center, Bartycka 18, 00-716 Warsaw, Poland} \\
     }

\abstract
{A quasar main sequence is widely believed to reveal itself through objects represented in a plane spanned by two parameters: the full-width at half-maximum (FWHM) of H$\beta{}$, and the ratio of Fe~{\sc ii} to H$\beta{}$ equivalent width. This sequence is related to the application to quasar properties of principal component analysis (PCA), which reveals that the main axis of variance (Eigenvector 1) is co-directional with a strong anti-correlation between these two measurements.} 
{We aim to determine whether the dominance of two Eigenvectors, originally discovered over two decades ago, is replicated in newer high-quality quasar samples. If so, we aim to test if a non-linear approach is an improvement on the linear PCA method by finding two new parameters which represent a more accurate projection of the variances than the Eigenvectors recovered from PCA.}
{We selected quasars from the XSHOOTER archive and a major quasar catalog to build high quality samples. These samples were tested with PCA.}  
{We find that the new high-quality samples do indeed have two dominant Eigenvectors as originally discovered. Subsequently we find that the fitting of a non-linear decay curve to the main sequence allows a new plane spanned by linearly independent axes to be defined, based on the distance along the decay curve as the main axis and the distance of each quasar data point from the curve as the secondary axis respectively.}
{The results show that it is possible to define a new plane based on the quasar main sequence which accounts for the majority of the variance. The most likely candidate for the new main axis is an anti-correlation with black hole mass. In this case the secondary axis likely represents luminosity. However, given the results of previous studies, inclination angle likely plays a role in H$\beta{}$ width.}

\authorrunning{Wildy, Czerny and Panda}
\titlerunning{Quasar Main Sequence: a line or a plane?}
\maketitle

\section{Introduction}

Active galactic nuclei (AGN) are complex objects. They contain a central black hole which accretes surrounding material, the inflowing matter having large angular momentum and forming an accretion disk. This disk is the dominant radiating component in the brightest active nuclei, which are known as quasars \citep{peterson97}. Although the accretion and radiation processes are complex, if we try to simplify the picture as much as possible we can reduce the number of parameters describing the active nucleus down to four: black hole mass, black hole spin, accretion rate (or, equivalently, the Eddington ratio), and viewing angle. In addition, we observe outflow phenomena (jets, winds) which are not necessarily uniquely determined by the parameters listed above and which affect considerably the observational properties of AGN. Also, AGN do not have to be well represented by a constant phenomenological type, since changing-look AGN \citep{matt03} have recently been catching a lot of attention . If such changing-look events are not just related to temporary obscuration, the picture may be much more complicated.

Nevertheless, attempts have been made to simplify the picture through determination of the most important parameters. In their pioneering paper in this field, \citet{boroson92} introduced Principal Component Analysis as a method for understanding AGN and showed that 13 measured quantities (line properties, multi-band indices, luminosity) can be combined into a sequence of Eigenvectors of decreasing importance. The two most important eigenvectors were able to explain 51 percent of the variance in these properties, obtained from a sample of 87 objects. Their principal Eigenvector (EV1) showed a strong anti-correlation between optical Fe~{\sc ii} equivalent width (EW) and both optical [O~{\sc iii}] EW and H$\beta{}$ FWHM. This research was continued later \citep{sulentic00,marziani01,sulentic07,kuraszkiewicz09} and lead to the appearance of the term Quasar Main Sequence, analogous to the well-established stellar main sequence. This analysis works only for brighter AGN like Narrow Line Seyfert 1 galaxies and quasars, and leaves out low luminosity AGN and blazars, but nevertheless it is an interesting step towards identification of a few key properties.

Following on from these papers, the study by \citet{shen14} claimed to have found a unified explanation for the EV1 behavior based on two parameters, namely the Eddington ratio and the quasar orientation. The plane of quasars on the FWHM(H$\beta{}$)--${\rm R_{Fe II}}$ plane, where ${\rm R_{Fe II}}$ is the ratio of the EW of Fe~{\sc ii} to the EW of H$\beta{}$, together with [O~{\sc iii}] EW, is illustrated in their Figure~1, showing the EV1 trends. It was theorized that increasing Eddington ratio drives both an increasing ${\rm R_{Fe II}}$ and decreasing [O~{\sc iii}] EW. The behavior of the H$\beta{}$ FWHM was explained as a combination of black hole mass and orientation angle influences.  However, this explanation is far from certain. For example, the measurement of Fe~{\sc ii} strength is subject to uncertainties due to the broadened and heavily blended nature of the emission. Also, subsequent papers have called into question the accuracy of the ${\rm R_{Fe II}}$ measurements \citep{sniegowska18} or have claimed the [O~{\sc iii}] EW to be subject to the influence of orientation \citep{risaliti11,bisogni17}. 

In this paper we apply the PCA technique to two high-quality quasar samples, using a different set of measured parameters than in previous studies. The aim is to check whether the minimum numbers of parameters (eigenvectors) needed to explain the majority of variance in the sample is a universal property of AGN, and thus can indeed be used to identify the minimum number of key parameters needed to describe an active nucleus in a typical Seyfert 1 galaxy or quasar.

\section{Using PCA for the analysis of quasar samples}

A PCA analysis works by initially finding the \emph{principal} axis along which the variance in the multi-dimensional space (corresponding to all recorded quasar properties) is maximized. This is known as \emph{Eigenvector 1}. Subsequent orthogonal eigenvectors, in order of decreasing variance along their respective directions, are found, until the entire parameter space is spanned. We analyze two different quasar samples using the PCA method with the aim of determining the relative importance of the eigenvectors, and we supplement them with the previous results obtained for the PG quasar sample by \citet{boroson92}. The key global parameters for these samples are given in Table~\ref{tab:zlumrange}.

\subsection{PCA analysis of an XSHOOTER quasar sample}

A high-quality sample of 30 quasar spectra, all located at similar redshifts ($z \sim 1.5$), was collected and analyzed in a series of three papers dedicated to understanding quasar spectral energy distributions and accretion disks \citep{capellupo15,mejarestrepo16,capellupo16}, hereafter referred to as the ``Capellupo sample''. These spectra were obtained using the \emph{XSHOOTER} instrument on the Very Large Telescope (VLT), allowing observation of several important rest-frame ultraviolet (UV) and optical lines at the selected redshift. These sources cover a relatively broad range of measured black hole masses (8.0~$\leq{}~log(M_{BH}/M_{\odot})~\leq{}$~9.5), Eddington ratios ($0.01~\leq{}~L/L_{EDD}~\leq{}~0.3$), inclination angles ($0.56~\leq{}~cos(i)~\leq{}~1.0$) and spin parameters ($0.5~\leq{}~a~\leq{~1.0}$). Given the broad ranges covered, this sample is excellent from the point of view of establishing the number and relative importance of whatever leading parameters are contributing to the Quasar Main Sequence. 

The 3-paper series presented a large number of spectral parameters measured from the Capellupo sample, we take 17 of these and use them for our PCA analysis. In the first attempt we use all 17 parameters, which are listed, along with abbreviated labels in brackets, as follows:

\begin{itemize}
\item FWHM of the H$\alpha{}$ line (fwhmha)
\item Dispersion of the H$\alpha{}$ line (sigha)
\item Rest-frame velocity offset of the H$\alpha{}$ line (dvha)
\item FWHM of the H$\beta{}$ line (fwhmhb)
\item Dispersion of the H$\beta{}$ line (sighb)
\item Rest-frame velocity offset of the H$\beta{}$ line (dvhb)
\item FWHM of the Mg~{\sc ii} $\lambda{}$2798 line (fwhmmg2)
\item Dispersion of the Mg~{\sc ii} $\lambda{}$2798 line (sigmg2)
\item Rest-frame velocity offset of the Mg~{\sc ii} $\lambda{}$2798 line (dvmg2)
\item FWHM of the C~{\sc iv} $\lambda{}$1549 line (fwhmc4)
\item Dispersion of the C~{\sc iv} $\lambda{}$1549 line (sigc4)
\item Rest-frame velocity offset of the C~{\sc iv} $\lambda{}$1549 line (dvc4)
\item Continuum luminosity at 3000~\AA{} (l3000)
\item Peak luminosity of C~{\sc iii}] $\lambda{}$1909 (lpc3)
\item Peak luminosity of C~{\sc iv} $\lambda{}$1549 (lpc4)
\item Peak luminosity of Si~{\sc iv}+O~{\sc iv}] near 1400~\AA{} (lpsi4)
\item Eddington ratio (lledd)
\end{itemize}

\noindent To do this, we use the \emph{prcomp} instruction in the \emph{R} statistical programming software.  We find that the first two eigenvectors, EV1 and EV2, have almost comparable contribution to the overall dispersion of the properties in the sample (see column headed ``Capellupo 17'' in Table~\ref{tab:EVparam}), with a considerable drop in the relative importance of the remaining eigenvectors. The velocity offsets were found to have a very low contribution to either EV1 or EV2, so the number of parameters was reduced down to 13 and the PCA analysis repeated. The contribution of EV1 and EV2 to the total variance increased in comparison to the previous case but their relative role remained practically the same (see column headed ``Capellupo 13'' in  Table~\ref{tab:EVparam}). 

\begin{figure}
 \centering
 \includegraphics[width=0.95\hsize]{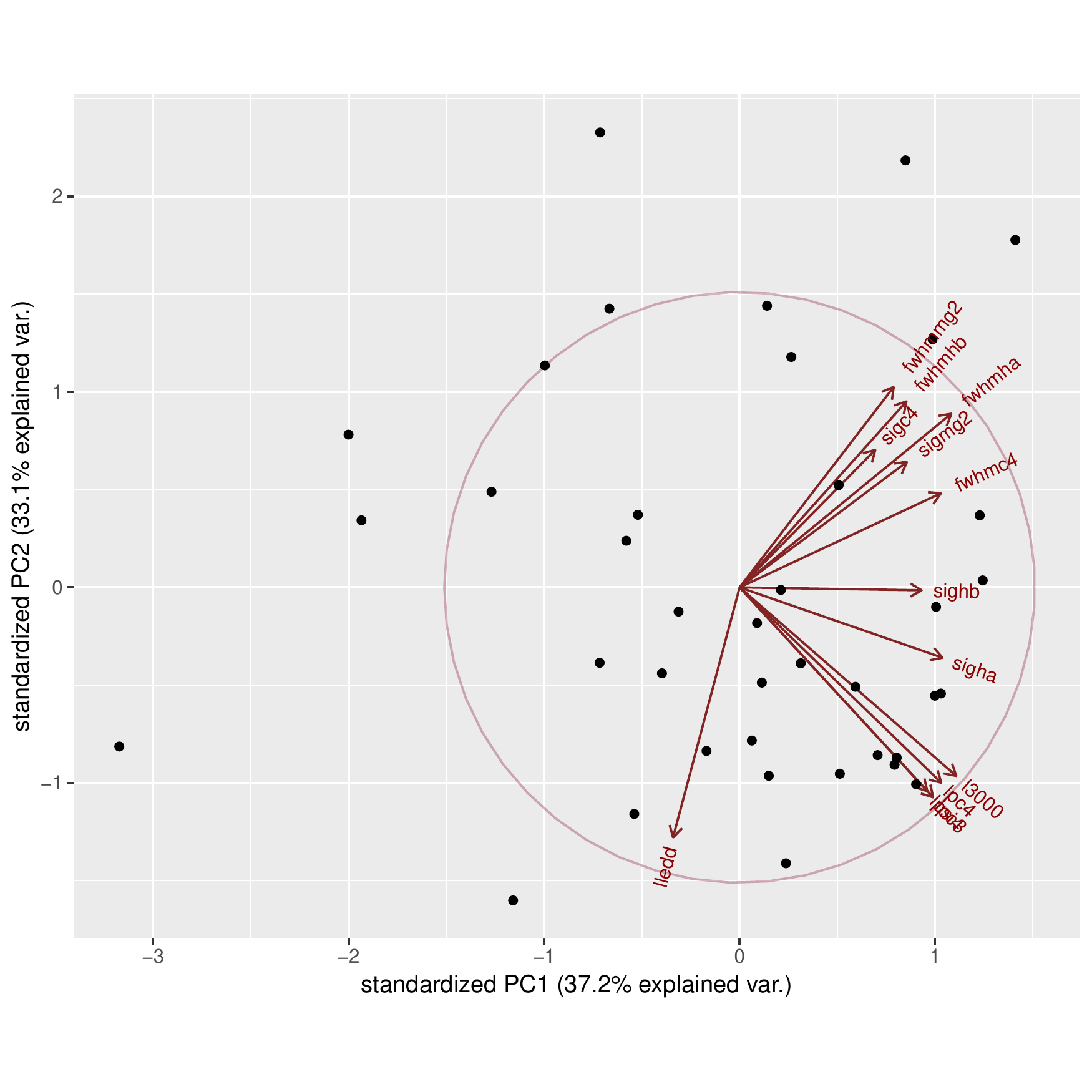}
 \caption{The graphical representation of the PCA decomposition of the Capellupo sample reduced to 13 parameters. Dots represent individual objects on the standardized EV1--EV2 plane, having values indicated by the axis labels. Arrows represent the loadings of the variables in the EV1--EV2 plane, with the unit circle included for scale reference.}
    \label{fig:capellupo}
\end{figure}

A graphical representation of the sample properties in the EV1--EV2 plane for the 13 Capellupo sample parameters is shown in Fig.~\ref{fig:capellupo}. We see that the Eddington ratio and the line width determined by the FWHM are almost aligned but with opposite orientation, which just graphically shows strong anti-correlation between the two quantities. On the other hand, the vectors representing the peak line luminosities and the continuum luminosity at 3000~\AA{} are perpendicular to the $EW - L/L_{Edd}$ trend. This visually illustrates that the EV1--EV2 plane is basically created by the luminosity and line kinematic width, or, equivalently, luminosity and the Eddington ratio.

This simple picture is perturbed by the orientations of the vector representing the line width measured by dispersion ($\sigma{}$) instead of FWHM. Indeed, the relation between $\sigma{}$ and FWHM is related to the line shape. If the line is well described with a single Gaussian, the ratio of $\sigma{}$ to FWHM is 0.426. However, if the line is of Lorentzian shape, the dispersion cannot be calculated and the ratio of the two quantities diverges. The study of \citet{collin06} showed that this ratio is partially affected by the viewing angle to the nucleus, and strongly related to the Eddington ratio. 

\subsection{PCA analysis of a high quality subsample of Shen et al. quasars}

We now describe a similar analysis using the general quasar catalog of \citet{shen11}. This catalog is built from the Sloan Digital Sky Survey (SDSS) and provides a number of measured quantities for quasars spanning a wide range of redshifts. We concentrate on eight choices for consistency with the previous analysis while maximizing the number of objects for which these properties are detectable within the SDSS spectral range. These quantities include seven used for the Capellupo sample (labels as described previously): l3000, fwhmha, lhb, fwhmhb, lmg2, fwhmmg2, lledd. In addition, the total luminosity in the H$\alpha{}$ line (lha) is included. The Shen catalog contains over 100\,000 quasars, however to ensure a high-quality sub-sample we select only those objects having seven of the eight quantities recorded as greater than five times the respective measurement error. The cataloged objects do not have an error measurement for the Eddington ratio, so this quantity was not used for the cut. The sample thereby reduced to 175 objects and is hereafter referred to in this paper as the ``reduced Shen'' sample. 

\begin{figure}
 \centering
 \includegraphics[width=0.95\hsize]{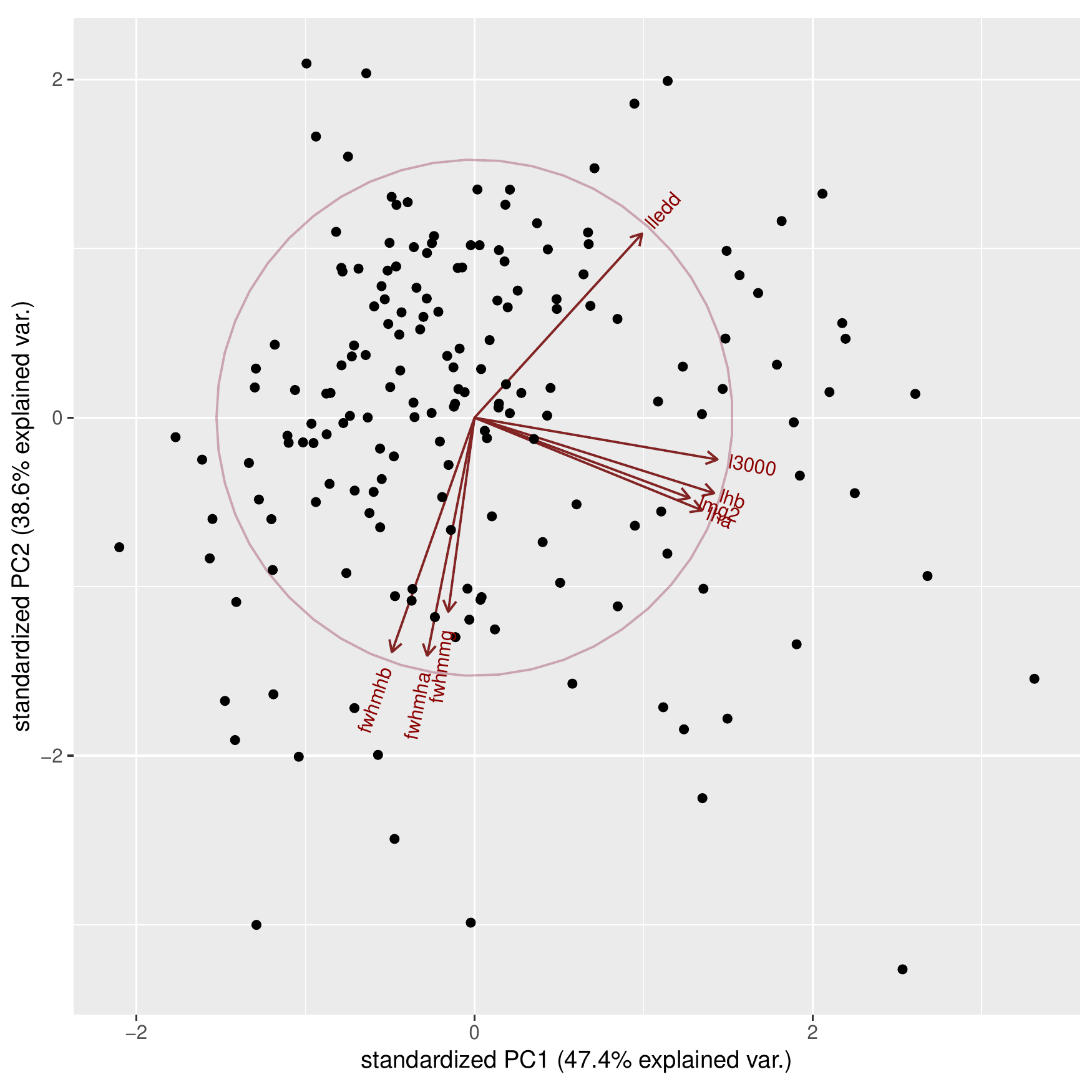}
 \caption{The graphical representation of the PCA decomposition of the Shen sample in the EV1--EV2 plane, analogous to Fig.~\ref{fig:capellupo}.}
    \label{fig:shen}
\end{figure}

The objects in the reduced Shen sample have average redshift z=0.37$\pm$0.02 and span a bolometric luminosity (${\rm L_{bol}}$) range of 45.00~$\leq$~log(${\rm L_{bol}/erg~s^{-1}}$)~$\leq$~46.76. Less than 10 percent (17 objects) are listed in \citet{shen11} as having a detection recorded in the FIRST catalog, hence the radio properties of this sample are not discussed further. A comparison of the redshift and luminosity ranges spanned by the Capellupo, reduced Shen and Boroson \& Green samples is provided in Table~\ref{tab:zlumrange}. A subsequent PCA analysis of this sample revealed again an almost equal role of the two leading eigenvectors (as seen in Table~\ref{tab:EVparam}), together responsible for 86.6 \% of the variability in the studied sample. The graphic representation of the EV1--EV2 plane in this case gives a remarkably simple picture, as seen in Fig.~\ref{fig:shen}. As before, the line width parameters are anti-correlated with the Eddington ratio to roughly form a single axis, while the parameters measuring luminosities are perpendicular to that axis, forming an independent direction in the plane.

\begin{table}
\caption{Redshift and luminosity ranges of each sample}
\label{tab:zlumrange}      
\centering                          
\begin{tabular}{l l l l}        
\hline\hline      
 & Capellupo & Reduced Shen & Boroson \& Green\\
\hline
Min z & 1.481 & 0.353 & 0.025\\
Max z & 1.646 & 0.390 & 0.472\\
Min log(${\rm L_{bol}/erg~s^{-1}}$) & 46.10 & 45.00 & 44.47\\
Max log(${\rm L_{bol}/erg~s^{-1}}$) & 47.28 & 46.76 & 46.87\\
\hline
\end{tabular} 
\end{table} 

\subsection{PCA results of Boroson \& Green for PG sample}

For completeness, we also list in Table~\ref{tab:EVparam} the results from the original paper by \citet{boroson92}. This sample contains 87 objects, for which 13 parameters were measured. We did not redo the analysis, but simply included the results given by \citet{boroson92} in their Table~4. In this sample the first two eigenvectors also dominate, although the effect is not as strong as in the reduced Shen sample. 

\begin{table}
\caption{Eigenvector contributions to the total variance in quasar samples}  
\label{tab:EVparam}      
\centering                          
\begin{tabular}{l r r r  r  r  r  r  r  r }        
\hline\hline      
EV    &  Capellupo & Capellupo & Reduced Shen & Boroson \& Green\\
      & 17 & 13   &  8  & 13 \\
\hline
1 &  29.4 \% &  37.2 \% & 47.4 \% & 29.2 \% \\
2 &  26.7 \% &  33.1 \% & 39.8 \% & 21.7 \% \\
3 &   9.0 \% &   8.2 \% &  7.3 \% & 10.5 \% \\
4 &   7.4 \% &   6.8 \% &  2.2 \% &  6.8 \% \\
5 &   5.5 \% &   4.1 \% &  1.8 \% &  6.0 \% \\
\hline
\end{tabular} 
\end{table} 

\section{Non-linear analysis of the quasar main sequence}

\begin{figure}
 \centering
 \includegraphics[width=0.95\hsize]{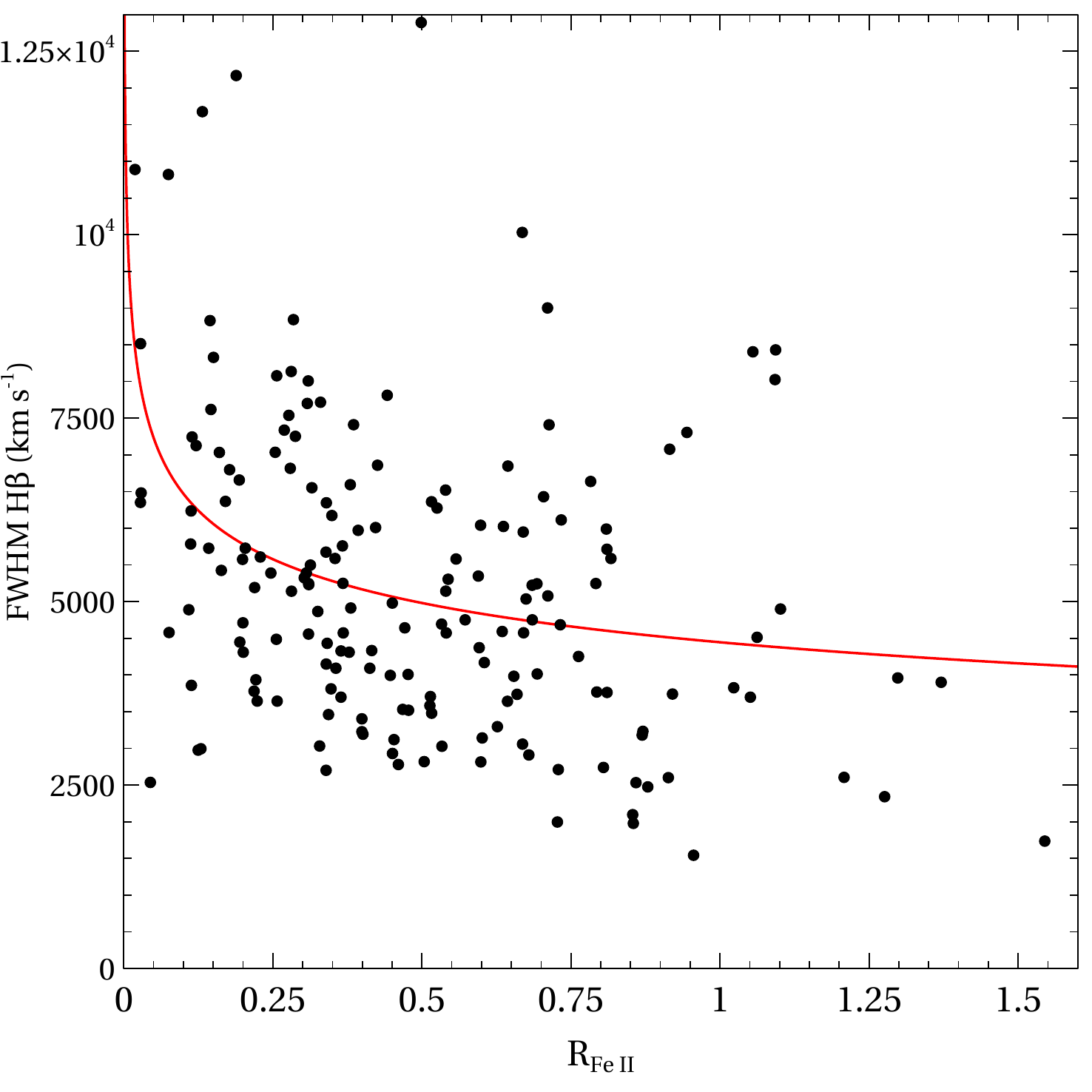}
 \caption{Position of reduced Shen quasars (black dots) on the FWHM(H$\beta{}$)--$\rm R_{Fe II}$ plane. The decay curve is in red. Blue diamonds mark the boundaries of unit distance intervals along the curve after the axes are normalized to their respective sample means, with the chosen zero point being first (uppermost left) and distance increasing towards the lower right.}
    \label{fig:qmsplot}
\end{figure}

The quasar main sequence appears as a linear structure in the PCA analysis, since this method only allows for a linear analysis of the measured quantities. On the other hand, a strong impression of a 1-D structure in the optical plane, defined by the H$\beta{}$ FWHM and ${\rm R_{Fe II}}$, is apparent \citep{sulentic00,marziani18}. There is an apparent steep decrease in H$\beta{}$ FWHM as a function of ${\rm R_{Fe II}}$ at low values of ${\rm R_{Fe II}}$, while a much shallower decrease is visible at high values. Following on from the results of the PCA analysis suggesting two dominant leading parameters, we attempt to define a new principal axis parameterized by a non-linear decreasing curve in the FWHM(H$\beta{}$)--${\rm R_{Fe II}}$ plane. As a test, we choose a simple decay curve having 2 free parameters of the form shown in Equation~\ref{eqn:nft} 

\begin{equation}
    y=\frac{1}{a+x^{b}}
    \label{eqn:nft}
\end{equation}

\noindent where $y$ and $x$ are the H$\beta{}$ FWHM and ${\rm R_{Fe II}}$ respectively, while $a$ and $b$ are the free parameters. To find $a$ and $b$ we use the \emph{nls} routine in \emph{R} to find a fit to the reduced Shen datapoints on the plane. The plane, together with the resultant non-linear fit, is shown in Fig.~\ref{fig:qmsplot}.

It is notable that the distribution of quasars in this high-quality sample looks distinctly different from the one in \citet{shen14} for the same plane. In that paper, the distribution is much more ``triangular'' (see their Figure 1). In our distribution, it is better approximated by the decay curve. The path along of the decay curve can be considered a non-linear analog of the EV1 derived from PCA, while the EV2 would correspond to the perpendicular offset from the curve. We therefore transform to a new co-ordinate plane parameterized by these new dimensions. This is achieved by first normalizing the H$\beta{}$ FWHM and ${\rm R_{Fe II}}$ to their respective sample means and then finding the nearest point on the curve to each quasar data point (the ``nearpoint''). Then, the two new parameters are assigned to each quasar data point.

The new principal parameter is defined by the distance along the curve, starting at the point where the mean-normalized H$\beta{}$ FWHM is 2.5 (corresponding to 13050~km~s$^{-1}$) and ending at each quasar nearpoint. This starting-point satisfies the requirement to have non-zero ${\rm R_{Fe II}}$, since the curve becomes infinite at ${\rm R_{Fe II}}$=0. It is also just greater than the maximum H$\beta{}$ FWHM in the quasar sample, and therefore is a suitable choice for the origin of the new main axis. The secondary parameter is therefore defined, for each quasar in the normalized plane, as the euclidean distance from the quasar datapoint to the corresponding nearpoint.

\begin{figure}
 \centering
 \includegraphics[width=0.95\hsize]{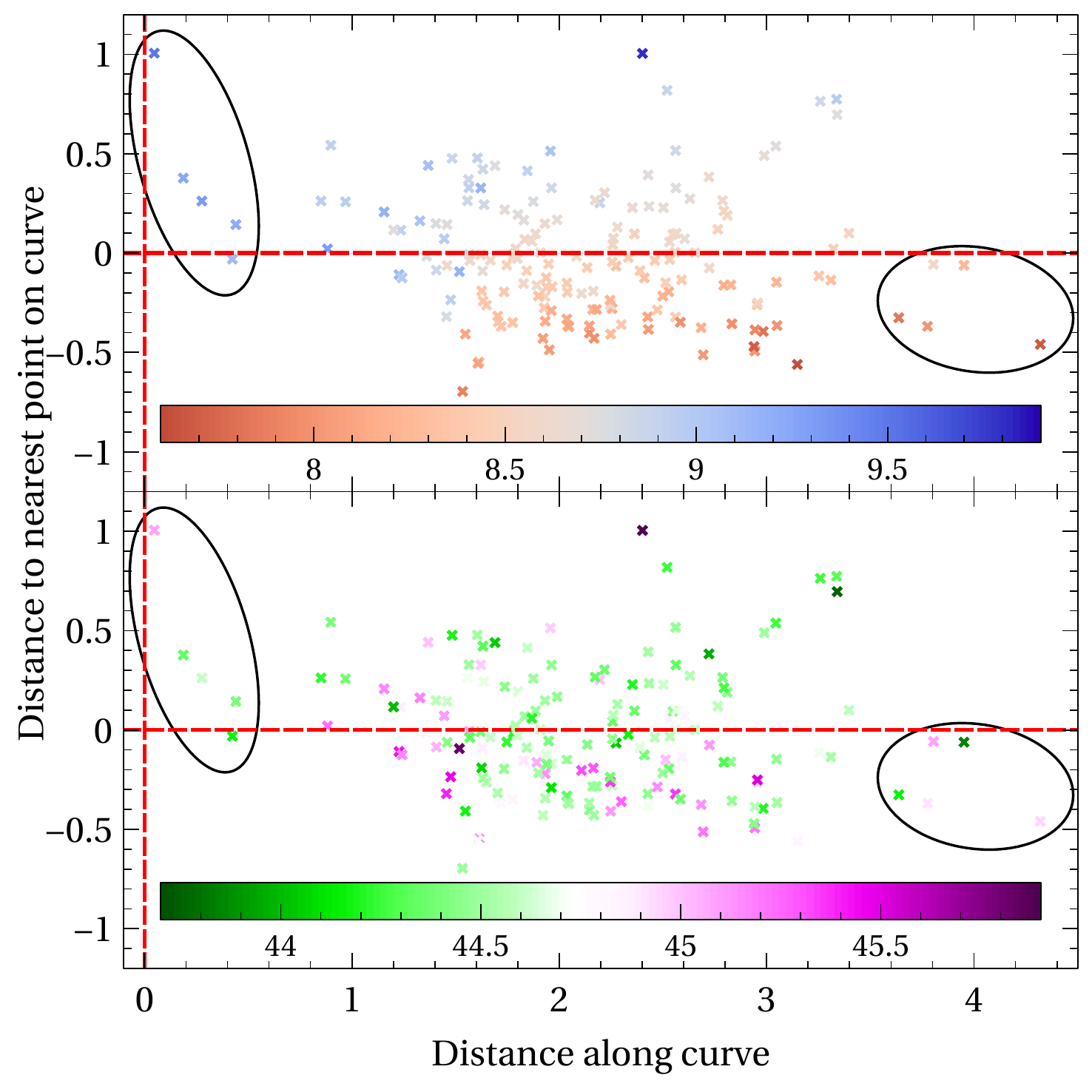}
 \caption{Two panels showing the positions of the reduced Shen quasars (colored crosses) on the new plane, defined as indicated on the axis labels. Two groups of five outlying quasars are inside the ellipses at either end of the horizontal axis. Top panel crosses are color-coded by black hole masses obtained from the Shen catalog, while bottom panel crosses are color-coded by 3000~\AA{} luminosity also obtained from Shen. Black hole mass units are log(M$_{\odot}$) while luminosity units are log(erg~cm$^{-2}$~s$^{-1}$~\AA{}$^{-1}$).}
    \label{fig:newplane}
\end{figure}

 If these new parameters do indeed represent a plane analogous to EV1 and EV2 from PCA, then they should be linearly independent. However, a Spearman rank correlation test reveals a significant anti-correlation (probability of no correlation p = 0.03) between the new parameters across the sample. A visual examination of the plane in Fig.~\ref{fig:qmsplot} suggests two outlying populations of five quasars at either end of the decay curve could be responsible. One group is those objects with FWHM(H$\beta{}$)$>$10\,500~km~s$^{-1}$, another is those objects with ${\rm R_{Fe II}}>$1.2. With these outliers removed, the significant correlation disappears. The new plane is illustrated in Fig.~\ref{fig:newplane}, with the removed outliers highlighted. The five blue diamond markers in Fig.~\ref{fig:qmsplot} correspond, in order from upper left to lower right, to the five indicated numerical values (0 to 4 respectively) on the horizontal axis of Fig.~\ref{fig:newplane}. 

\section{Discussion}

It is known from reverberation mapping \citep{blandford82} that AGN black hole mass scales with the H$\beta{}$ FWHM, since the calculated masses (subject to the virial coefficient) are found to follow the M--$\sigma{}$ relation determined for quiescent galaxies \citep{woo10}. This means Eddington ratio should correlate negatively with H$\beta{}$ FWHM, and indeed a significant anti-correlation between the two is found using a Spearman correlation test on the reduced Shen sample. Is it therefore possible that the new principal parameter is measuring black hole mass?  It would certainly be required that larger values of the new principal parameter indicate lower masses, given that this is the direction of lower H$\beta{}$ FWHM. However it is also found that Eddington ratio tends to increase along the new principal parameter in the reduced sample. Therefore if this parameter is measuring black hole mass, it would require a decline in Eddington ratio at higher mass, possibly caused by the existence of a sub-Eddington limit \citep{steinhardt10,garofalo19}.

The conclusion that the black hole mass is the basic parameter changing along the quasar main sequence was formulated recently by \citet{fraixburnet17}. They performed an analysis of two samples of 215 and 85 low-z quasars (z~$\leq{}$~0.7), using a machine learning approach in the form of an unsupervised multivariate classiffication technique. They included the measurement of seven independent parameters. This approach combines objects into trees according to the similarity of the parameters, and finally it can suggest an evolutionary pattern. The conclusion from their analysis was that the black hole mass, which grows over the cosmological time, allows a reconstruction of the evolutionary pattern across the optical plane (see their Fig. 7). They interpret the effective coupling between mass and Eddington ratio as caused predominantly by the study of the local Universe, where sources having both large black hole mass and high Eddington ratio are not present. This, in turn, is the result of the galaxy evolution. Large black hole masses are hosted by large mass galaxies, as implied by the black hole mass - bulge mass correlation \citep{magorrian1998,denicola2019}, and the gas content is relatively low with respect to the stellar mass in large galaxies. This last trend is already present at redshift $\sim 2$, but it deepens with the redshift decrease, when the star formation rate and the gas content strongly decrease \citep[see e.g.][and the references therein]{peeples2011,popping2015,decarli2019}. Thus, the coupling is effectively due to limited resources of material in galaxies hosting increasingly massive black holes..

However, it is not entirely clear how lower black hole mass (equivalently higher Eddington ratio in this case) could give rise to greater ${\rm R_{Fe II}}$. A lower mass should produce a harder ionizing continuum, which would favor H$\beta{}$ production over Fe~{\sc ii}. A possible explanation for this is given in \citet{marziani18} based on different accretion disk modes. In this case when the Eddington ratio is high, the accretion disk behaves as a slim disk \citep{abramowicz88}, with ``puffed up'' walls close to the black hole. Conversely, at low Eddington ratio, the disk behaves more like a traditional thin disk \citep{shakura73}. If this picture is true, low mass objects are likely to produce a flat, shielded region where Fe~{\sc ii} is preferentially formed over H$\beta{}$, solving the puzzle.

What if the new secondary parameter in this case were luminosity? A correlation test indicates that this axis is strongly anti-correlated with luminosity, which is in accordance with the broad line region (BLR) radius-luminosity relationship determined in \citet{kaspi05}. The slope of the decay curve indicates that the luminosity would have greater projection into ${\rm R_{Fe II}}$ as compared with H$\beta{}$ FWHM at higher masses than at lower masses. This may be related to the difference in accretion disk mode, since at high masses, differing luminosities could vary ${\rm R_{Fe II}}$ through ionization effects without getting close to Eddington luminosity due to the sub-Eddington limit, while at low masses ionization effects are countered by the tendency of higher luminosity to create a greater shielded region in the slim disk regime.

It is likely that the H$\beta{}$ FWHM is, to some degree, positively correlated with inclination angle, given the findings of studies which use radio data as an orientation indicator \citep{wills95,aars05}. This supports the idea of BLR H$\beta{}$ emission originating in a flattened region, since at higher inclination the orbital velocity will have greater projection into the plane of the line-of-sight. The study of \citet{zamfir08} showed that sources having Fanaroff-Riley type II (FR II) morphology are more common at higher H$\beta{}$ FWHM than those sources which are radio core-dominated. The FR II quasars are more likely to be viewed at high inclination as their radio lobes project into the plane of the sky as seen from Earth, further supporting the relationship between H$\beta{}$ width and inclination.

It is also possible that ${\rm R_{Fe II}}$ declines with increasing orientation as Fe~{\sc ii} is a relatively low ionization line and therefore may form in an even more flattened configuration than H$\beta{}$, since it necessitates shielding from the ionizing continuum. Indeed, evidence for this scheme is hinted at in \citet{bisogni17}, where Fe~{\sc ii} strength seems to drop to zero at very high [O~{\sc iii}] equivalent width (and hence inclination), while H$\beta{}$ also drops but remains nonzero. Given the shape of the decay curve in Fig~\ref{fig:qmsplot}, it is therefore conceivable that the new main parameter anti-correlates with inclination angle. However even in such a case it is likely that mass plays a strong role, as objects at low values of the new parameter predominantly have large line widths, which can only be achieved by high mass objects irrespective of inclination. It seems unlikely that inclination could form the main driver of the new secondary parameter, since increasing inclination should tend to increase the width of H$\beta{}$ while suppressing the ${\rm R_{Fe II}}$ ratio (as discussed previously). An examination of Fig.~\ref{fig:qmsplot} shows this is impossible, as all normal lines to the best-fit curve describe a change of both parameters in the same sense (positive or negative).

Another possibility is the involvement of black hole spin, however, for the BLR, this should mainly affect the hardness of the incident continuum, with higher spin resulting in a harder continuum. According to \citet{panda18}, based on photoionization simulations object-to-object variations in continuum peak temperature cannot explain the quasar main sequence. We can therefore say that it is most likely that the new principal parameter represents an anti-correlation with black hole mass, with the new secondary parameter representing an anti-correlation with luminosity. However, this does not rule out inclination having a separate influence on both H$\beta{}$ FWHM and ${\rm R_{Fe II}}$.

\subsection{Outlying points}

\begin{figure}
 \centering
 \includegraphics[width=0.95\hsize]{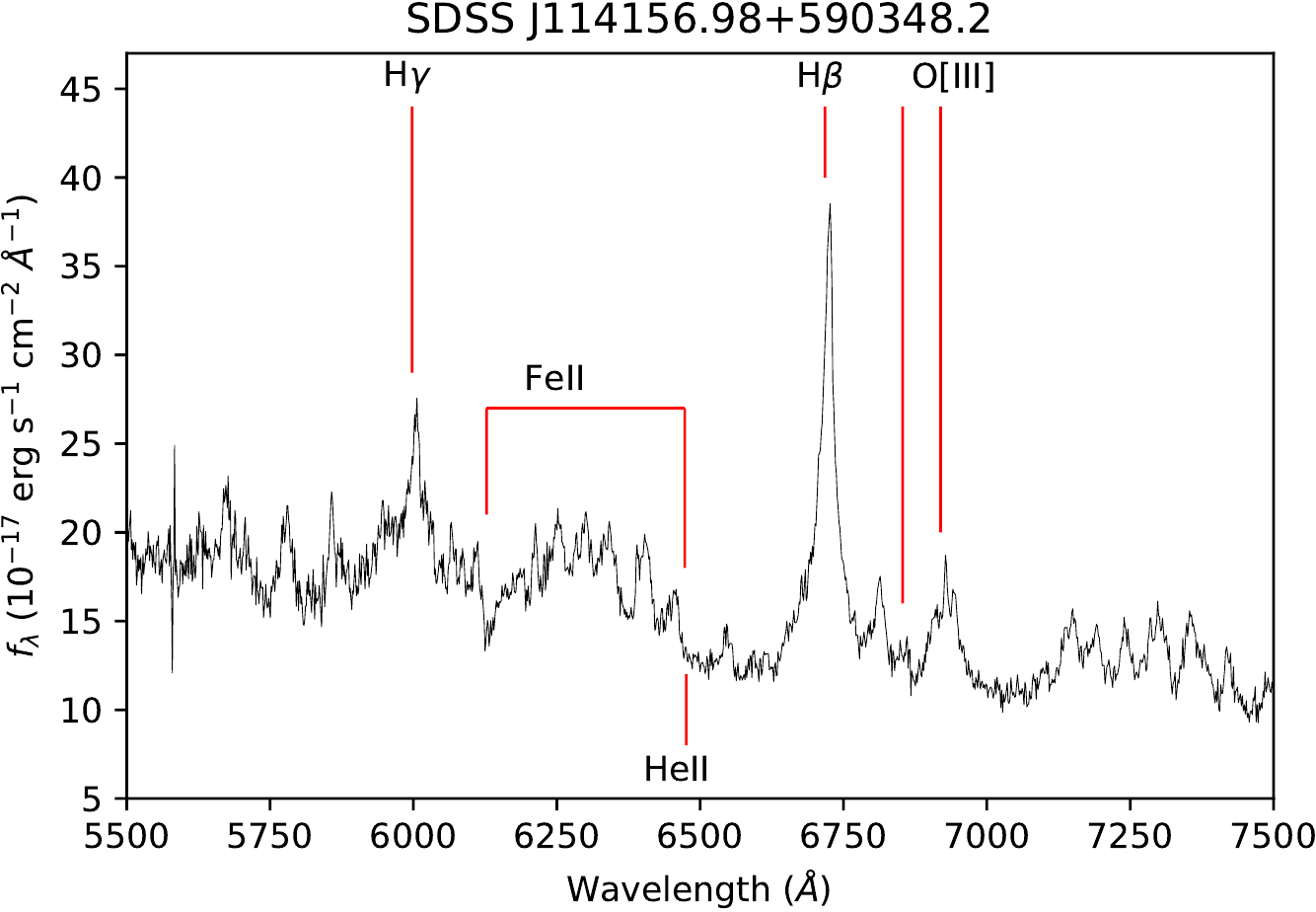}
 \caption{One of the outliers in the right lower corner Fig.~\ref{fig:newplane}: this quasar is an extreme representative of NLS1 class, with narrow Balmer lines and strong Fe~{\sc ii} emission, but otherwise it does not show any peculiarities in the spectrum.}
    \label{fig:outlier6}
\end{figure}

As can be seen in Fig.~\ref{fig:newplane}, the outlying points form two groups of five at either end of the new principal axis. It is worth considering the possible reasons these outliers exist. Studies of the quasar main sequence tend to show that the bulk of quasars are concentrated in the region 0$\leq$${\rm R_{Fe II}}$$\leq$1.0. This is also the case for our sample (see Fig.~\ref{fig:qmsplot}). It is not certain that values of ${\rm R_{Fe II}}$ greater than approximately 1.3 as derived from the \citet{shen11} catalog are accurate, since a more detailed investigation into a sample of such objects by \citet{sniegowska18} indicated that a large majority (21/27) did not in fact have such a high value. This raises the possibility that the five outliers having the highest value of the new principal axis may be further from the bulk of quasars than warranted due to an overestimation of ${\rm R_{Fe II}}$. An example spectrum of one of these outliers is shown in Fig.~\ref{fig:outlier6}.

Any mis-calculation of R$\rm _{Fe~II}$ or H${\beta}$ FWHM in \citet{shen11} would result from inaccurate modeling of one or both of the Fe~{\sc ii} and H~$\beta{}$ emission. Accurate determination of the strength and shape of the emission profiles strongly depends on the correct placement of the disk continuum, which is difficult in a spectral region heavily contaminated by line emission. Also, the Fe~{\sc ii} blend spans approximately 4340~\AA{} to 4680~\AA{}, which overlaps with the broad He~{\sc ii} emission at 4686~\AA{}. It thus may be difficult to distinguish between the two and result in He~{\sc ii} flux in this region being wrongly attributed to Fe~{\sc ii} or vice-versa. In cases of very large width the H$\beta{}$ line itself could overlap with the Fe~{\sc ii} blend as seen, for example, in Figure 6 of \citet{marziani12}, resulting in confusion between the two. 

\begin{figure}
 \centering
 \includegraphics[width=0.95\hsize]{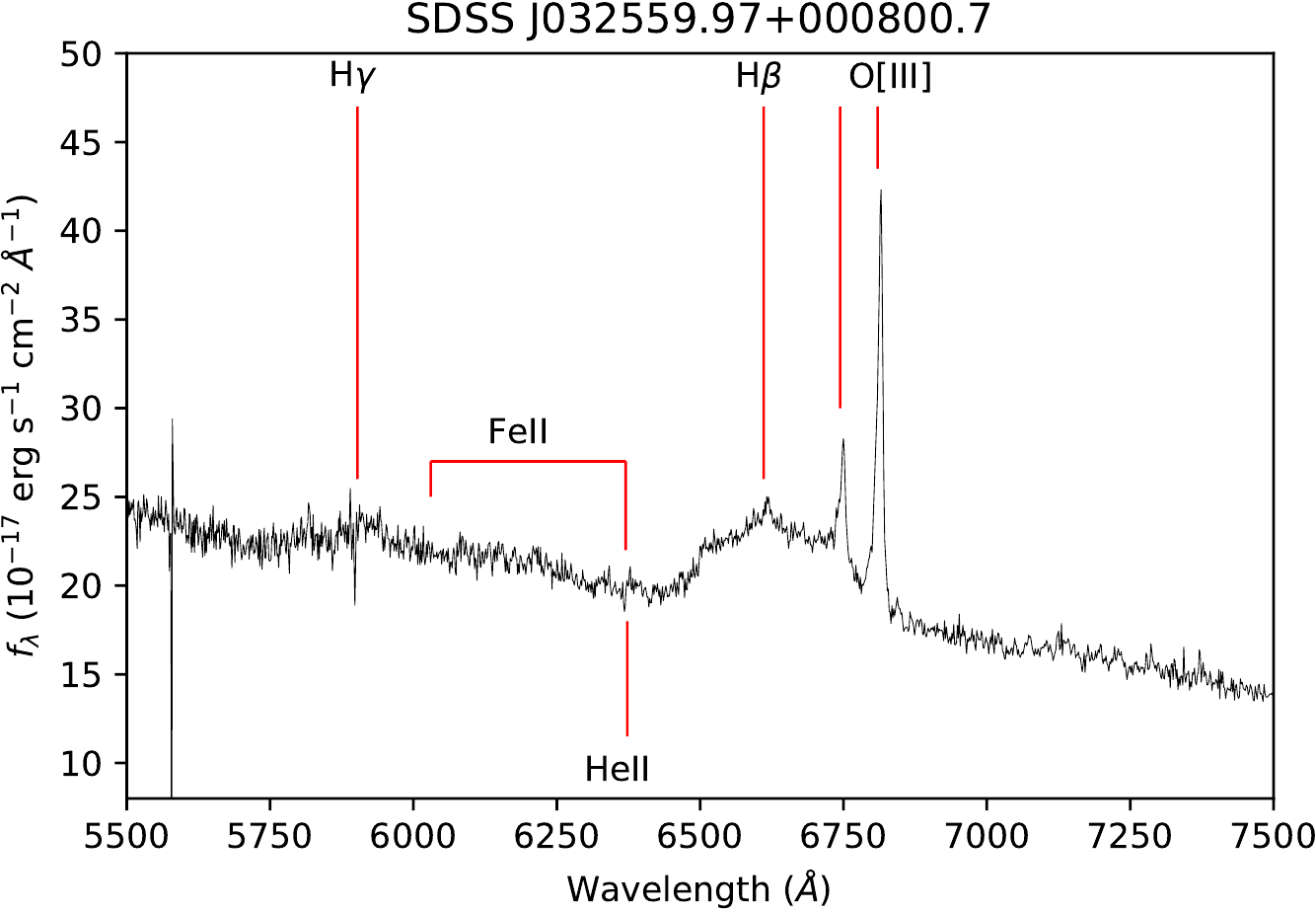}
 \caption{The extreme outlier in the left upper corner of Fig.~\ref{fig:newplane}: this quasar and neighboring outliers are characterized by the double-peak profile of the broad component of H$\beta$ line.}
    \label{fig:outlier1}
\end{figure}

For the group having low values of the new principal axis, it is possible that the true H$\beta{}$ kinematic line width is overestimated, pushing them away from the quasar bulk. This may result from variable asymmetric broad emission line caused by complexities such as shocks generating emission in addition to photoionization \citep{shapovalova10}. From examination of the H$\beta{}$ profiles in these objects, this explanation is indeed plausible, as they appear to have a characteristic double-peaked profile. An example spectrum where this phenomenon is visible is shown in Fig.~\ref{fig:outlier1}.

\begin{figure}
 \centering
 \includegraphics[width=0.95\hsize]{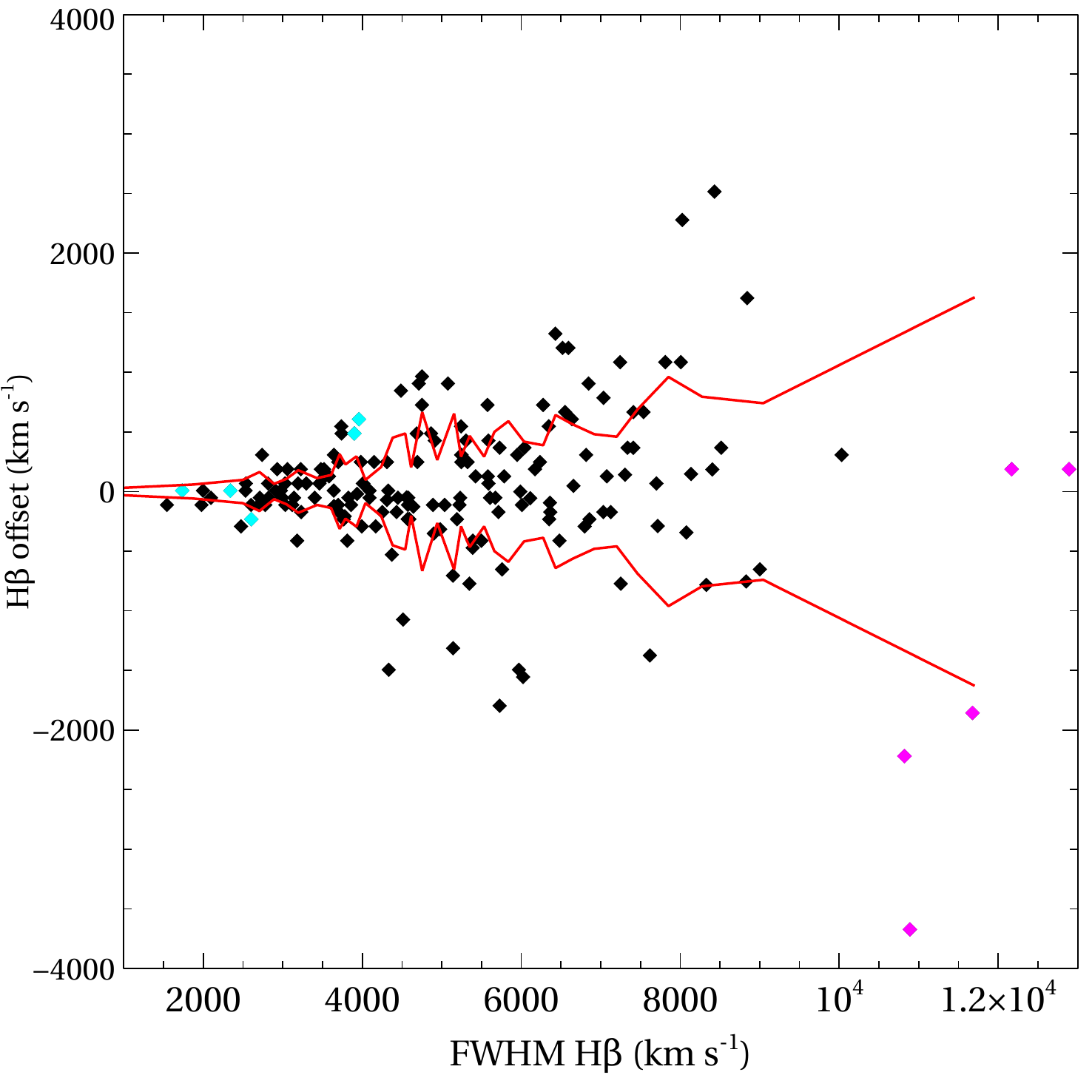}
 \caption{Widths and H$\beta{}$ offsets of the reduced Shen quasars (black diamonds). Outlying points are coloured differently (cyan for low principal axis values at magenta for high principal axis values. Red lines indicate an envelope calculated from a five-point moving magnitude average.}
    \label{fig:offs}
\end{figure}

The offset of the broad component centroid of H$\beta{}$ from the theoretical quasar rest-frame wavelength can also give clues about the nature of the outlying points. Asymmetries can result in large offsets, and these asymmetries can either be intrinsic or an artifact of incorrectly modelled profiles. In Fig.~\ref{fig:offs} the H$\beta{}$ widths and offsets are listed for all quasars in the reduced Shen sample, with both sets of outliers highlighted. As can be seen, larger offsets generally result from broader lines. The fraction of non-outlying points with offsets placing them outside the mean envelop (61/165 or 37 percent) is smaller than that of outlying points (6/10 or 60 percent). This could indicate that one of the previously mentioned mechanisms may be responsible for the outliers. The object with an offset of $-$3670~km~s$^{-1}$ appears extreme in this plot and therefore is of note, however upon inspection the spectrum did not appear unusual.

\section{Conclusions}

We have shown, for a high quality quasar sample, that it is possible to describe the behaviour of the quasar main sequence in the FWHM(H$\beta{}$)--${\rm R_{Fe II}}$ plane using two linearly independent parameters, as suggested by the PCA analysis. One of these is distance along a decay curve, which accounts for the majority of variance, and the other is the distance in parameter space of each object from the curve. The distribution of quasar data points in the FWHM(H$\beta{}$)--${\rm R_{Fe II}}$ plane is distinctly different from that of \citet{shen14}, which is approximately triangular. 

If the two new parameters dominate the behavior of the quasar main sequence, then we can suggest what they may be. Given the already well-known positive relationship between H$\beta{}$ width and black hole mass, together with the  additional possibility of ${\rm R_{Fe II}}$ dependence on accretion disk mode and shielding effects, it seems very plausible that the distance along the decay curve represents the direction of decreasing black hole mass. In this case the secondary axes would probably represent luminosity. The effect of inclination on H$\beta{}$ FWHM however is strongly suggested by previous studies, so it probably has an influence, even if it is not one of the primary drivers of the new parameters.

\begin{acknowledgements}
The project was partially supported by Polish grant No. 2015/17/B/ST9/03436/.

\end{acknowledgements} 
\bibliography{bib_pca}

\bibliographystyle{aa}

\end{document}